\documentclass[%
 aip,
 amsmath,amssymb,
 reprint,%
]{revtex4-1}

\usepackage{graphicx}
\usepackage{dcolumn}
\usepackage{bm}

\usepackage[utf8]{inputenc}
\usepackage[T1]{fontenc}
\usepackage{mathptmx}

\draft 

\begin{document}


\title{Shape induced segregation and anomalous particle transport under 
spherical confinement} 



\author{Abhinendra Singh}\thanks{equal contribution}
\affiliation{Pritzker School of Molecular Engineering, University of Chicago, Chicago, Illinois 60637, USA}
\affiliation{James Franck Institute, University of Chicago, Chicago, Illinois 60637, USA}
\author{Jiyuan Li}\thanks{equal contribution}
\affiliation{Pritzker School of Molecular Engineering, University of Chicago, Chicago, Illinois 60637, USA}
\author{Xikai Jiang}
\affiliation{State Key Laboratory of Nonlinear Mechanics, Institute of Mechanics, Chinese Academy of Sciences, Beijing 100190, China}
\affiliation{Pritzker School of Molecular Engineering, University of Chicago, Chicago, Illinois 60637, USA}
\author{Juan P. Hern\'andez-Ortiz}
\affiliation{Department of Materials and Nanotechnology, Universidad Nacional de Colombia--Medell\'in, Medell\'in, Colombia, 050034}
\affiliation{Colombia/Wisconsin One-Health Consortium, Universidad Nacional de Colombia--Medell\'in, Medell\'in, Colombia, 050034}
\email{jphernandezo@unal.edu.co}
\author{Heinrich M. Jaeger}
\affiliation{James Franck Institute, University of Chicago, Chicago, Illinois 60637, USA}
\affiliation{Department of Physics, The University of Chicago, Chicago, Illinois 60637, USA}
\author{Juan J. de Pablo}
\affiliation{Pritzker School of Molecular Engineering, University of Chicago, Chicago, Illinois 60637, USA}
\affiliation{Materials Science Division, Argonne National Laboratory, Lemont, Illinois 60439, USA}
 \email{depablo@uchicago.edu}


\date{\today}

\begin{abstract}
Colloid or nanoparticle mobility under confinement is of central importance to a wide range of physical and biological processes. Here, we introduce a minimal model of particles in a hydrodynamic continuum to examine how particle shape and concentration affect the transport of particles in spherical confinement. Specifically, an immersed boundary--General geometry Ewald-like approach is adopted to simulate the
dynamics of spheres and cylinders under the influence of short- and long-range fluctuating hydrodynamic interactions with appropriate non-slip conditions at the confining walls. An efficient $O(N)$
parallel finite element algorithm is used, thereby allowing simulations at high concentrations, while a Chebyshev polynomial approximation is implemented in order to satisfy the fluctuation--dissipation theorem.  A concentration--dependent anomalous diffusion is observed for suspended particles.
It is found that introducing cylinders in a background of spheres, i.e. particles with a simple degree of anisotropy, has a pronounced influence on the structure and dynamics of the particles. First, increasing the fraction of cylinders induces a particle segregation effect, where spheres are pushed towards the wall and cylinders remain near the center of the cavity. This segregation leads to a lower mobility for the spheres relative to that encountered in a system of pure spheres at the same volume fraction. Second, the diffusive-to-anomalous transition and the degree of anomaly -- quantified by the power law exponent in the mean square displacement vs. time relation -- both increase as the fraction of cylinders becomes larger. These findings are of relevance for studies of diffusion in the cytoplasm, where proteins exhibit a distribution of size and shapes that could lead to some of the effects identified in the simulations reported here. 
\end{abstract}


\maketitle 

\section{Introduction}
Colloidal and nanoparticle diffusion in confined environments arises in a wide range of scientific and engineering systems, including living cells, mesoporous materials, or microfluidic devices  
~\cite{happel2012low, koch2011collective, hudson2006nanotechnology, richert1996geometrical,seymour2004anomalous}. It is also of interest for energy generation processes that rely on salinity or electrostatic gradients in pores~\cite{gao2014high, zhang2017ultrathin,darling2018perspective,waldman2018janus}.
 In the particular case of the cytoplasm, the diffusion of biomolecules underpins a variety of intracellular metabolic, translational and locomotion processes, to name a few.~\cite{minton1981ap, fulton1982crowded, Konopka:2006kf, Ellis:2001bu, Ellis:2001gy}.
Interestingly, particle diffusion in these confined systems is often found to be severely hindered and anomalous~\cite{selhuber2009variety, weiss2004anomalous, hofling2013anomalous}.
The mechanisms behind those observations, however, remain poorly understood.

Several literature studies have examined particle mobility in living cells~\cite{bicout1996stochastic, McGuffee:2010bi} by relying on Brownian dynamics (BD) simulations. In such studies, biological macromolecules have been represented as spheres, and numerical simulations have found evidence of hindered diffusion, in agreement with experimental results.
Majority of previous studies, however, have failed to consider hydrodynamic interactions between particles or between particles and the confining walls.
Some exceptions are provided by the work of Ando et al.~\cite{Ando:2010hg} and Chow et al.~\cite{Chow:2015ja}, who included hydrodynamic interactions between particles, but did not enforce the no-slip boundary condition at the walls. More recently, Stokesian dynamics (SD) simulations of spheres by Aponte-Rivera et al.~\cite{zia:2016, zia2018} considered both far-- and near--field (lubrication)
hydrodynamic interactions (HI) between particles and walls.
The authors demonstrated that HI have a pronounced influence on the local structure and the short--time and long--time diffusive behavior of particle suspensions. The framework employed by these authors relied on SD, and was restricted to a homogeneous system of spherical particles~\cite{zia2018}.

Recently, we have introduced an efficient computational approach in order to overcome some of the limitations of other available numerical approaches for hydrodynamic interactions.
In particular, this approach can be easily extended to particles of arbitrary shape dispersed in a confined geometry also of arbitrary
shape~\cite{Zhao:2017, kounovsky2017electrostatic, unpublishedkeyA}.
An Immersed--Boundary (IB) method is used to represent the suspended finite--sized particles.
A parallel Finite Element General geometry Ewald-like method (pFE-GgEm)~\cite{Zhao:2017} is used to calculate the confined Green's functions, which relies on a Chebyshev polynomial approximation to satisfy the fluctuation-dissipation theorem.
In recent work~\citep{unpublishedkeyA}, we relied on this approach to compare the structure of pure spherical and pure cylindrical particles confined in a spherical cavity.
It was found that cylindrical  particles diffuse slower as compared to spherical particles of the same volume and at the same volume fraction, and that for cylinders the transition from the diffusive to the sub-diffusive regime occurs at a lower volume fraction.

The studies mentioned above focused on pure spheres or cylinders confined in a spherical cavity. The more relevant case of mixtures of spheres and cylinders was not considered.
Indeed, in applications (e.g. cytoplasm or colloidal suspensions) one rarely deals with systems of pure mono-disperse spheres, and it is therefore of interest to consider how mixtures behave relative to their pure counterparts. Note that limited experimental evidence with mixtures of particles of different sizes and shapes indicates that cells exhibit preferential accumulation of some particles near the nucleus~\cite{kodali2007cell, kolhar2012polymer}. In those cases, size based segregation was explained on the basis of a ``sieving effect'' that has been advanced in the dry granular segregation literature~\cite{savage1988particle, gray2018particle}.
An explanation for shape-based segregation was not proposed in that work.
Other experimental work, including a study of centrifugation of colloidal rods and spheres~\cite{sharma2009shape}, and a study in which milli--meter sized glass beads and rods were subject to strong vibration \cite{caulkin2010geometric}, have also reported segregation effects based on particle shape, and proposed that hydrodynamic forces based on the different shapes contribute to that segregation.

Our particular goal here is to provide a standard against which past and future observations of segregation and diffusion can be compared by simulating mixtures of particles of equal volume but having a spherical or a cylindrical aspect ratio $h_c/r_c =2$. By doing so, we seek to rationalize past reports with new evidence for size-based segregation and mobility gradients in systems where dimensions and interactions are perfectly controlled, thereby eliminating or avoiding some of the complexity that arises in laboratory experiments. The outline of this paper is as follows: in Section~\ref{sec:model} we describe our numerical setup and methodology.
Our results on the structure and dynamics of mixtures of spheres and cylinders are presented in Section~\ref{sec:results}. We conclude the manuscript with a discussion of our findings in Section~\ref{sec:conc}, along with a possible outlook for future studies.

\section{Model and method}\label{sec:model}
The system considered here consists of $N$ semi--rigid particles embedded in a viscous fluid of viscosity $\eta$ that are enclosed in a spherical cavity of radius $R$.
The equations of motion under the condition of zero Reynolds number and zero Stokes number are given by
\begin{equation} \label{eqn-solids}
\mathbf{F}^H + \mathbf{F}^B + \mathbf{F}^{C} + \mathbf{F}^{EV} + \mathbf{F}^{ext} = 0~,
\end{equation}
where $\mathbf{F}^{H}$ is the $6N$ vector containing the hydrodynamic force/torque, $\mathbf{F}^{B}$ is the Brownian force/torque, $\mathbf{F}^{C}$ is the force/torque containing configuration terms,
$\mathbf{F}^{EV}$ represents force/torque excluded volume contributions and $ \mathbf{F}^{ext}$ includes any external force/torque.

Evolution of the suspended particles, using Eqn.~\eqref{eqn-solids}, is achieved using the grand mobility or resistance tensors that relate the hydrodynamic force/torque with the translational and rotational velocities of the particles~\cite{pozrikidis,ladyz,power_book}.
Approaches like SD~\cite{Brady1988,SierouBrady01,brady2010} and boundary integral methods (BIM)~\cite{pozrikidis,hernandez_book} are used extensively to
solve the ``mobility problem".
The regularized Stokeslets~\cite{cortez01}, the accelerated BIM~\cite{Kumar:2012ev} and the Immersed  Boundary (IB)~\cite{Peskin2002,Atzberger:2007p3828,kallemov2016immersed,fiore2017rapid,sprinkle2019brownian} 
provide examples of numerical methods
developed to improve computational efficiency by simplifying or avoiding the calculation
of the single- and double-layer hydrodynamic potentials of suspended particles.
On the case of the Immersed Boundary (IB) approach, the surfaces of the suspended solids are represented by a distribution of discrete force densities on a surface mesh ($N_\text{IB}$ immersed boundary nodes) that, together with a surface force description and Stokes equations, leads to the evolution of the suspended particles. This is the approach that we use in this work.

The probability distribution function for the surface mesh positions in a Lagrangian frame of reference evolves according to a convection-diffusion equation of the Fokker-Planck type~\cite{risken-book}.
We assume a continuous probability density and use the Chapman-Kolmogorov equation white noise to obtain an equivalent stochastic differential equation for the motion of the mesh points~\cite{Ottinger1996}
\begin{equation} \label{eq:eqn2}
d\mathbf{R} = \left[
						\mathbf{U_{0}}+\mathbf{M}\cdot\mathbf{F}+
			            \frac {\partial  } {\partial \mathbf{R} } \cdot \mathbf{D}
			            \right] dt
			            + \sqrt{2} \mathbf{B} \cdot d\mathbf{W},
\end{equation}
where $\mathbf{U_{0}}$ denotes a $3(N\times N_\text{IB})$ vector of the unperturbed fluid velocity generated by external pressure differences or shear at the mesh point positions; $\mathbf{M}$ is the mobility tensor that includes the Stokes' drag and the pair-wise Stokeslets accounting for the hydrodynamic interactions between mesh points; $\mathbf{D}=k_BT\mathbf{M}$ is the $(3N\times N_\text{IB})\times (3N\times N_\text{IB})$ diffusion tensor; $\mathbf{F}$ is a $3(N\times N_\text{IB})$ vector of the non-Brownian and non-hydrodynamic forces;
$k_B$ is the Boltzmann constant; $T$ is the temperature;
$\mathbf{M} \cdot \mathbf{F}$ is a convection term that represents the bead velocities arising from hydrodynamic interactions; the divergence of the diffusion tensor, $ {\partial}/{\partial \mathbf{R}} \cdot \mathbf{D} $, is the first diffusive term resulting from the configuration-dependent mobility of the confined mesh points; $d\mathbf{W}$ is a random vector, the components of which are obtained from a real-valued Gaussian distribution with zero mean and variance $dt$, and it is coupled to the diffusion tensor through the fluctuation-dissipation theorem, $ \mathbf{D}=\mathbf{B} \cdot \mathbf{B}^{T} $; and finally, the second diffusive term, $ \sqrt{2} \mathbf{B} \cdot d\mathbf{W}$, represents the Brownian displacement that results from collisions between mesh points and the surrounding (implicit) solvent.

The main challenge in simulating a stochastic process using Eqn.~\eqref{eq:eqn2} is the fact that the mobility tensor, $\mathbf{M}$, cannot be constructed explicitly under confinement for arbitrary geometries.
This implies that the fluctuating velocity, $\mathbf{U}$, the divergence of the diffusion tensor, $\nabla\cdot\mathbf{D}$, and the diffusion tensor decomposition, $\mathbf{B}$, must be implemented in a way such that the scheme is matrix-free.
To address this issue,
we have developed an efficient $O(N)$ numerical algorithm, parallel Finite Element - General Geometry Ewald-like Method (pFE-GgEm)\cite{Zhao:2017}. The algorithm uses
(i) the General geometry Ewald-like method (GgEm)\cite{HernandezOrtiz:2007} for a matrix-free product of the mobility tensor with any vector, $\mathbf{M}\cdot\mathbf{F}$; (ii) a mid-point algorithm, proposed by Fixman\cite{Fixman:1986}, that avoids the explicit calculation of $\nabla\cdot\mathbf{D}$; and (iii) a Chebyshev polynomial approximation for the $\mathbf{B}\cdot d\mathbf{W}$ product that uses GgEm to avoid the explicit calculation of $\mathbf{D}$.  The algorithm is able to handle arbitrarily shaped confining walls.

Each particle is represented by a discretized surface, whose details are available in Ref.~\cite{unpublishedkeyA}. Using the Immersed Boundary (IB) method~\cite{Pranay2010}, the force distributions at these particles are discretized as distributions of regularized point-forces.
In particular,
\begin{equation} \label{eq:eqn3}
\boldsymbol{\rho}_\text{IB}^{f}(\mathbf{x})=\sum_{\nu=1}^{N_\text{IB}}
\mathbf{f}^\text{C}_{\nu} \delta_\text{IB}(\mathbf{x}-\mathbf{x}_{\nu}),
\end{equation}
where $\mathbf{f}^\text{C}_\nu$ represents the constitutive force acting on $\nu$-th surface node (point force with an excluded volume of radius $a$), $N_\text{IB}$ represents the number of surface nodes that are used to represent the suspended finite-size particles, $\delta_\text{IB}$ is the modified Gaussian regularization function. The regularization parameter $\xi_\text{IB}$ in $\delta_\text{IB}$ is related to the characteristic length $h$ for the node spacing on the particle surface, i.e. $\xi_\text{IB}\sim h^{-1}\sim a^{-1}$.
By doing this, we ensure that the regularized force on each node is spread over the length scale of the associated surface elements to prevent fluid from "penetrating" the particles.

The volume of spheres and cylinders is the same. Each surface node is linked to the neighboring node as well as to the center-of-mass point of the particle using an elastic spring with stiffness $k$.
A spring network is formed for every particle, which results in an internal nodal force that resists deformation and maintains its shape. At the same time, a repulsive Lennard-Jones (LJ) potential is used for particle-particle and particle-wall excluded volume interactions. The ratio between mesh and particle size controls the number of surface nodes on each particle. Increasing the number of nodes improves accuracy but also increases the computational cost.  In a previous study~\cite{unpublishedkeyA}, we showed that a spring stiffness $k=200$ is sufficient to simulate ``semi-rigid" particles, where despite the high concentration of particles, excluded volume interactions do not alter the particle shape. In addition, we found that particle discretization at the level of $N_{IB} = 20$ is enough to avoid fluid penetration, satisfy Stokes' law and provide
the correct diffusional behavior.  In this work, however, we use spheres and cylinders discretized with $N_{IB} = 88$ to ensure extremely high accuracy.

\begin{figure} 
\centering
\includegraphics[width=0.95\columnwidth]{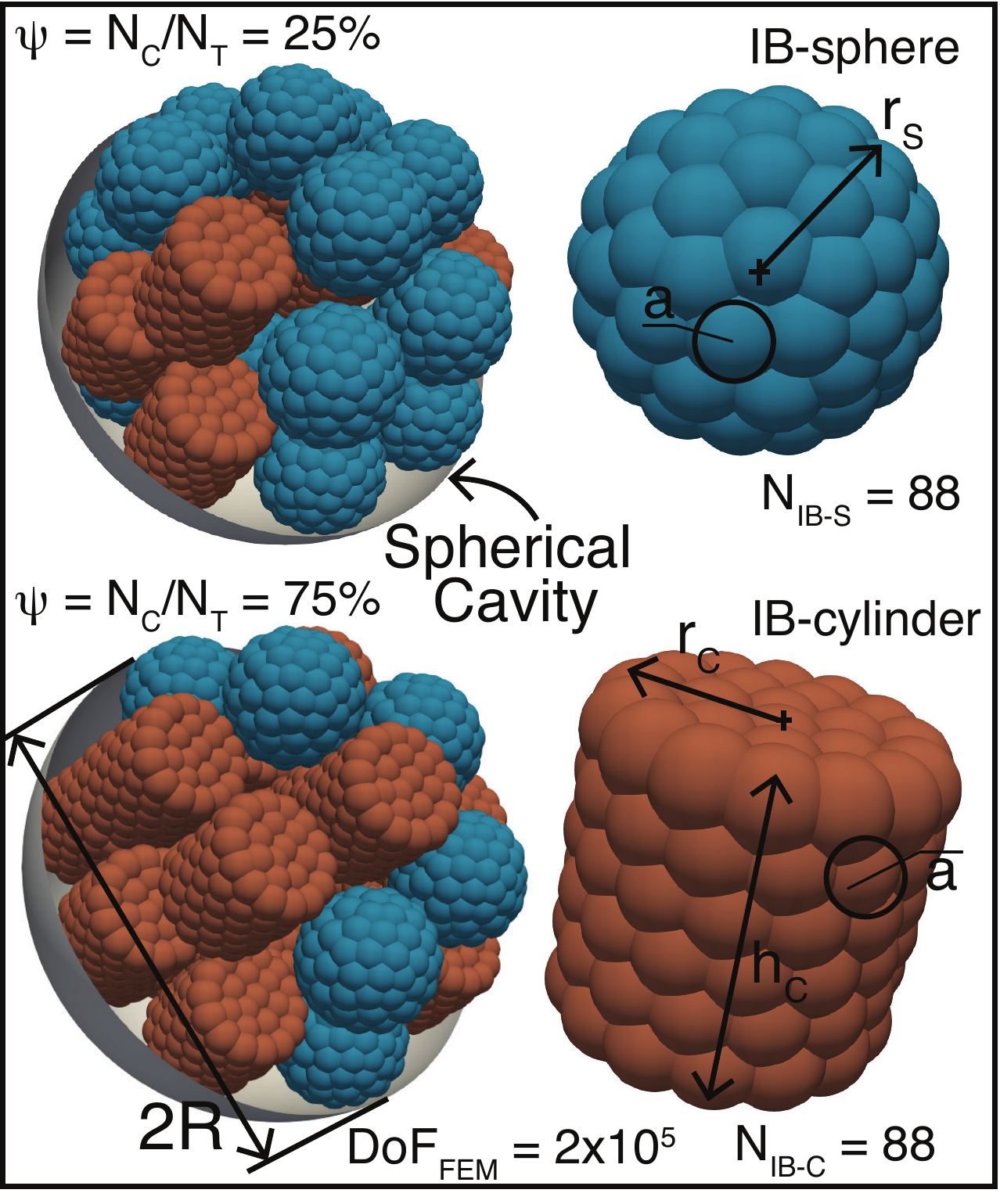}
\caption{Snapshots of the spherical cavity of radius $R$ containing spherical and cylindrical particles with $\phi_\text{HI} = 0.2$
for fraction of cylinders $\mathrm{\psi} = \text{N}_\text{C}/\text{N}_\text{T}$ being 0.25 and 0.75.
The spherical particles radius is $r_S$, while the size of the cylindrical particles is determined by $r_C$ and $h_C$.
The surface of the particles is given by a collection of discrete nodes that are connected to six neighbors, similar
to boundary element discretizations, and with a characteristic node separation of $a\sim h \sim \xi_\text{IB}^{-1}$.
A repulsive Lennard-Jones excluded volume is included on each surface node, shown
schematically in the particles' cross section by the black circles. The characteristic size of the repulsion is
given by $\sigma = 2.2a$.}
\label{fig:system_representation}
\end{figure}

In what follows, the characteristic units are: $a$ for length, $a^2\zeta/k_BT$ for time, $k_BT$ for the energy and $k_BT/a$ for the force. $\zeta$, the node friction coefficient is related to the fluid viscosity $\eta$ and $a$ through Stokes' law, i.e., $\zeta=6\pi\eta a$, and the unit diffusivity, $D_0$, is defined as the diffusivity of a sphere in an infinite fluid with viscosity $\eta$, i.e., $D_0 = k_BT/6\pi\eta a$.

\begin{figure*}[!ht]
\centering
\includegraphics[width=0.95\textwidth]{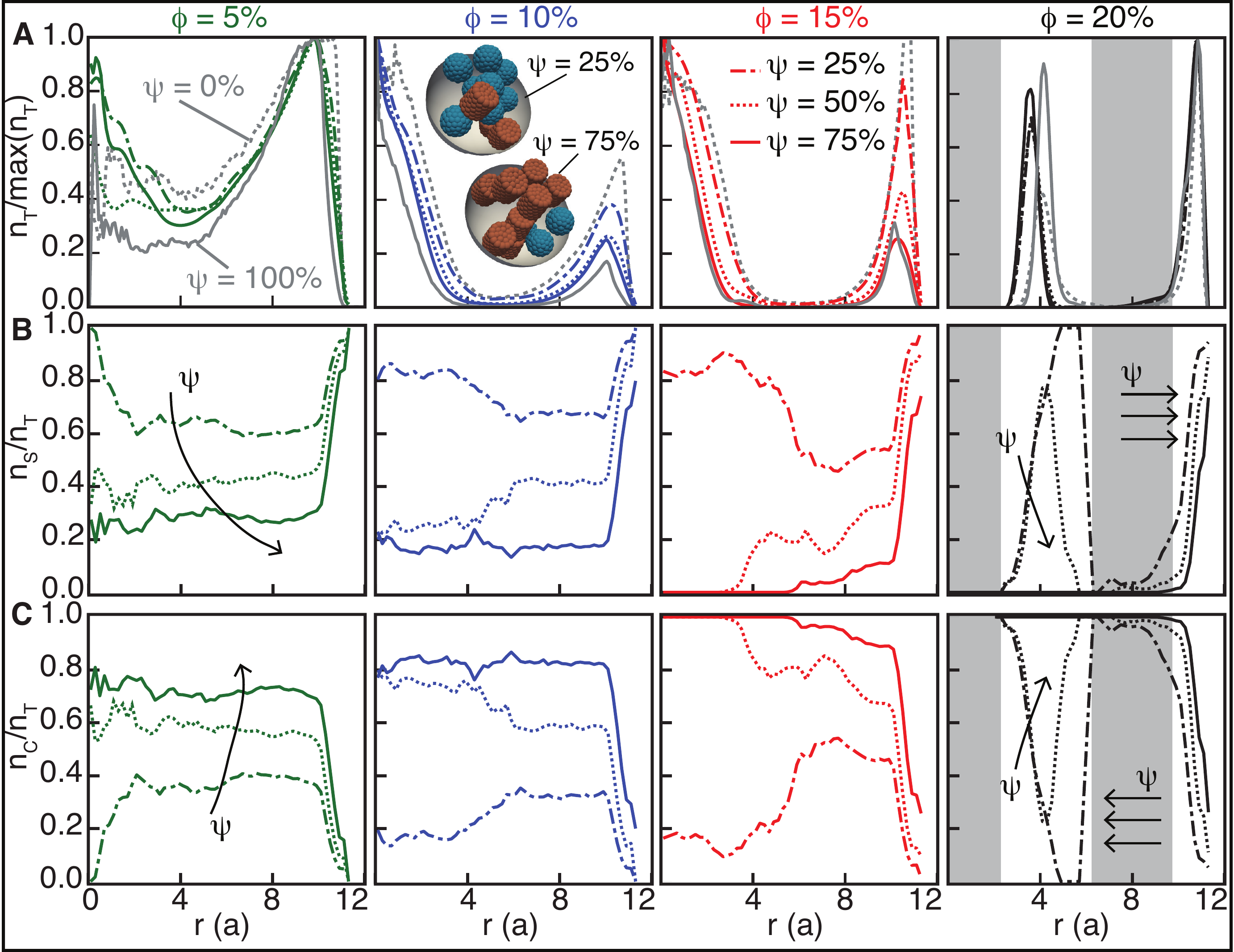}
\caption{
Particle number density in the mixture of spherical and cylindrical particles within a spherical cavity of radius $R=15$ as a function of radial distance.
The radii of spheres and cylinders are $r_S=3$ and $r_C=2.62$, respectively. The cylinder has an aspect ratio of 2, i.e., $h_C=2r_C=5.24$.
(A): Number density of all particles scaled with the maximum density $\text{n}_\text{T}/\text{max}(\text{n}_\text{T})$ for particle
concentrations $\phi=$ 5\%, 10\%, 15\%, and 20\% (from left to right).
 For each particle concentration, different fractions of cylinders ${\psi}$ are displayed along with the pure sphere (0\%) and pure cylinder (100\%) cases.
(B): Relative number density of only spherical particles scaled with the total density $\text{n}_\text{S}/\text{n}_\text{T}$ for various particle
concentrations $\phi=$ 5\%, 10\%, 15\%, and 20\% (from left to right).
(C): Relative number density of only cylindrical particles scaled with the total density $\text{n}_\text{C}/\text{n}_\text{T}$ for various particle
concentrations $\phi=$ 5\%, 10\%, 15\%, and 20\% (from left to right).
}
\label{fig:density}
\end{figure*}

\section{Results}\label{sec:results}
We consider different particle mixtures of spheres and cylinders
suspended in a Newtonian viscous fluid within a spherical cavity of radius $R=15$.
The spherical particle has a radius $r_s=3$, leading to a hydrodynamic volume of $V_{\mathrm {HI}} = 4/3\pi r_s^3$.
The cylinders have an aspect ratio of 2, i.e., $h_c = 2r_c$,  where $r_c=2.62$ is the radius and $h_c$ is the height.
Figure~\ref{fig:system_representation} shows several details of our simulations and representative snapshots for $\phi=0.2$ with different cylinder fractions.

In our semi-rigid particle model, there are two ways to define the particle concentration in a cavity of volume $V$.
A hydrodynamic volume fraction can be defined as $\phi_\text{HI} = N_T V_\text{HI}/V$;
a second one is based on the excluded volume, $\phi_\text{EV} = N_T V_\text{EV}/V$, where for spheres and cylinders we have $V_\text{EV} = 4/3\pi (r_S+a)^3$ and $V_\text{EV}=\pi (r_C+a)^2(h_C+2a)$,  respectively.
In the remainder of the article, we will use the hydrodynamic volume fraction $\phi_{\mathrm {HI}}$ (referred to as $\phi$ in the rest of the paper)
to denote the concentration of the particles.
In this work, we explore $\phi = [5\%, 10\%, 15\%, 20\%]$;
this would correspond to $\phi_{\mathrm{EV}}$ = [12\%, 24\%, 36\%, 48\%].

\subsection{Structure of sphere and cylinder mixture.}
We begin by analyzing the structure of mixtures through the local particle number density. To calculate it, the spherical cavity is discretized into $m$ evenly-spaced spherical shells along the radial direction, leading to a shell radius of the $i\text{-th}$ shell that is given by $b_i = (i+0.5)R/m$.
The particle number density is then given by $n(r_i)=\langle N(r_i)/V_i \rangle$, where $N(r_i)$ is the number of particles in the $i-th$ shell with volume $V_i$, and is at a distance $r_i$ from the center of the cavity; $\langle \rangle$ represents the ensemble average over time.
We calculate the number density for all particles, only spheres, and only cylinders, and denote them by $n_T$, $n_S$, and $n_C$, respectively.

Figure~\ref{fig:density} displays the number density for particles within the cavity for various particle concentrations $\phi$ and different fractions of cylinders for each $\phi$.
Figure~\ref{fig:density}A shows the number density for all particles $n_T (r)$ within the cavity for various particle concentrations $\phi$ and different fractions of cylinders $\psi$.
Cases with  $\psi=$ 0\% and 100\% cylinders refer to packings with pure spheres and cylinders, respectively.
The density of particles is scaled with the maximum number density for each case.
A common observation is that the scaled density profiles exihibit a peak close to the wall, decreases in the bulk and then increases at the  center of the cavity.
For low concentration, $\phi = 5\%$, the peak in scaled number density near the wall is independent of $\psi$, while it decreases with $\psi$ in the bulk. At particle concentration $\phi = 10\%$, the scaled density shows a peak at the center, decreasing with increasing $r$ and increasing again near the wall.
A similar observation can also be drawn for the case of $\phi = 15\%$.
Note that for the two cases ($\phi= 10, 15\%$) the scaled density at the center is higher than that near the wall. Another common feature of these two cases is that the difference between the two scaled densities decreases with increasing $\psi$, implying that the addition of cylinders enhances the heterogeneity in the local density. At the highest concentration considered here $\phi = 20\%$, we observe a layered structure with two distinct peaks at $r=3$ and 10, along with a depletion zone in the regions $r<2$ and $5<r<7$.
The peak position of the layered structure for the pure cylinder case is slightly different compared to other fractions, and the difference between the two peaks decreases with increasing $\psi$.

To further understand the local particle density, we analyze the relative density of spheres and cylinders.
Figure~\ref{fig:density} (middle row) displays the number density of spheres relative to the total density as a function of $r$ for various values of $\phi$ and $\psi$ = 25, 50, and 75\% for each case.
We observe that the scaled sphere density relative to the total number density $n_S/n_T$ is highest close to the wall and decreases with increasing fraction of cylinders.
$n_S/n_T$ decreases with increasing fraction of cylinders in the bulk and is always greater than zero for low particle concentrations $\phi=5\%$ and 10\%.
$n_S/n_T$ becomes zero for higher particle concentrations $\phi = 15\%$ and 20\% at large cylinder fraction (75\%); only cylinders are found in this range of $r$, as confirmed by $n_C/n_T$ being equal to $1$, as shown in Fig.~\ref{fig:density} (bottom row).
These two observations demonstrate that only cylinders are present in the interior of the cavity and that  spheres are close to the wall.
Another point to note is that the numerical values of the scaled densities for $\psi=$ 25\% and 50\% in the bulk are more "separated" compared to the differences between  $\psi=$ 50\% and 75\%.

These results serve to establish the equilibrium segregation of spheres to the walls induced by a subtle difference in particle shape but for the same particle volume.
As a side note, we mention here that the difference in particle volume may not be the only reason for the observed segregation in experiments~\cite{kolhar2012polymer}.

\begin{figure}[!htb]
\centering
\includegraphics[width=0.95\columnwidth]{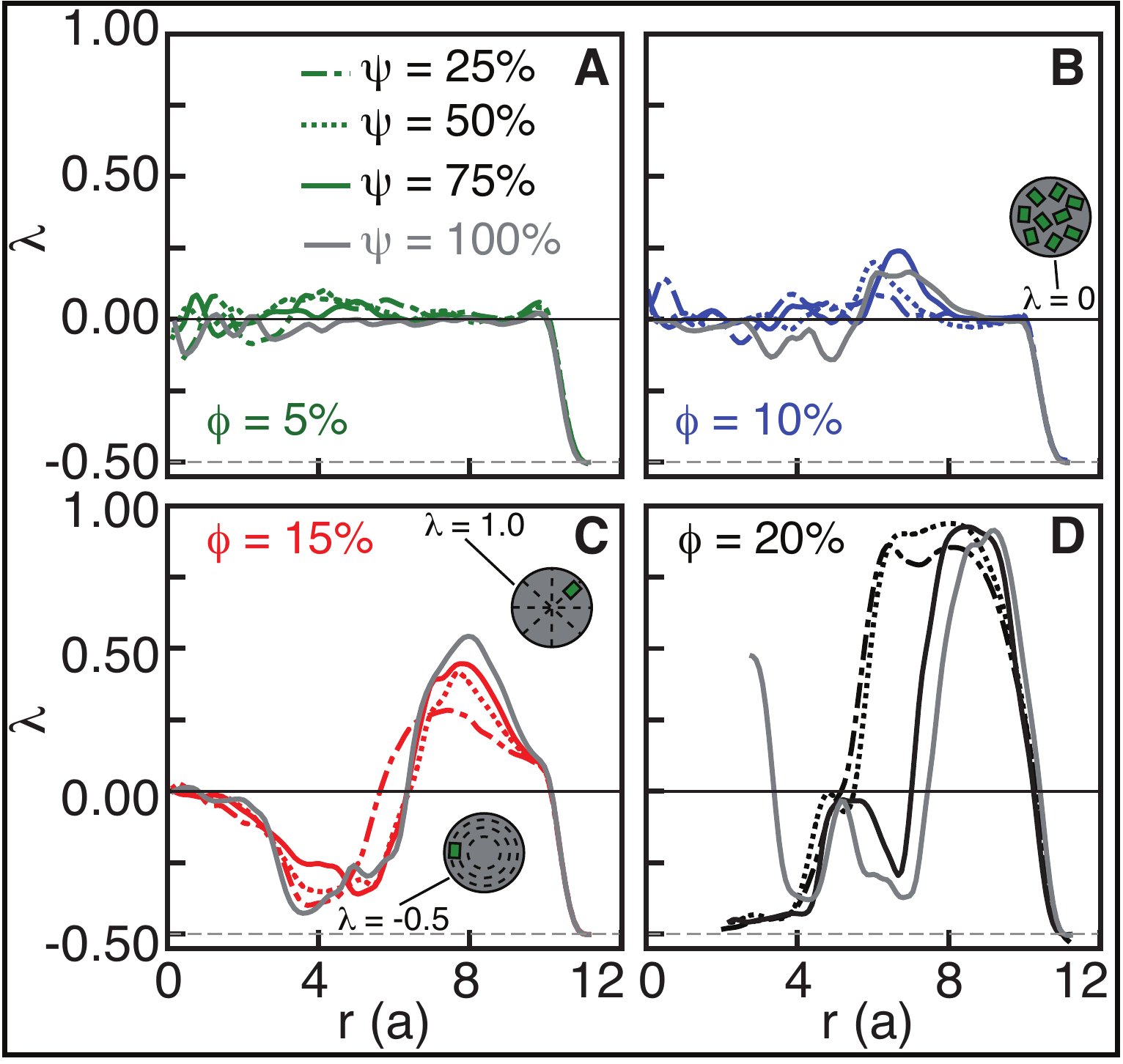}
 \caption{
 Orientational order parameter $\lambda$ of cylindrical particles within a spherical cavity of $R=15$
 as a function of radial distance for particle concentration $\phi=$ (A) 5\%, (B) 10\%, (C) 15\%, and (D) 20\%.
The radius of cylinders are $r_C=2.62$ and the height $h_C=2r_C=5.24$. }
 \label{fig-lambda_cylinder}
\end{figure}

Next, we analyze the orientational order parameter for different particle concentrations and cylinder fractions.
The orientational order parameter is defined as $\lambda=\frac{1}{2}\langle3\cos ^2 \theta-1\rangle$,
where $\cos\theta={\mathbf{m}\cdot\mathbf{n}}/({||\mathbf{m}||\cdot||\mathbf{n}||})$, $\mathbf{m}$ is the vector parallel to the centerline of the cylinder and $\mathbf{n}$ is the vector connecting the cavity center and the cylinder's center-of-mass. A parameter
$\lambda$ is often used in liquid crystalline systems to quantify the nematic ordering~\cite{lebwohl1972nematic, martinez2017};
$\lambda=0$ corresponds to a random/disordered configuration, whereas $\lambda$ is unity for ordered morphologies, with the cylinder axis being coaxial with the radial direction of the spherical cavity (radial phase), and $\lambda = -1/2$ when all cylinders are aligned transversal to the radial direction (concentric phase).
Figure~\ref{fig-lambda_cylinder} displays $\lambda$ for various particle concentrations with different fractions of cylinders. A common observation is that, very close to the wall, the order parameter is $\lambda=-1/2$ irrespective of the volume fraction, indicating a concentric phase close to the wall.
We also find that $\lambda$ fluctuates around zero in the bulk for low volume fractions, i.e, $\phi=5\%, 10\%$, indicating a disordered configuration of cylinders. 
For moderate concentrations, $\phi=15\%$, $\lambda$ is zero close to center and is negative with increasing $r$, reaching a minimum and increasing further with $r$ to reach a maximum value of 0.5; $\lambda$ then decreases with $r$ reaching -0.5 close to the wall.
For the highest concentration, we find that $\lambda=-1/2$ at both the center and close to the wall, and we also find a depletion zone with no particles for $r<2$.
For this concentration, we find another ordered state with $\lambda \sim 1$ in the region $7<r<10$.
The ordered morphology arises from segregation in the cavity.
At the highest volume fraction $\phi=20\%$, the cylinders display ordered morphologies, i.e., perpendicular to the radial direction very close to the wall and parallel to the radial direction for  $7<r<10$.
With increasing $\psi$, the cylinders push the spheres to the wall in order to minimize free volume and gain orientational order.

Excluded volume potential calculations yield
$10.88 k_B T$ for a single sphere, $2.3 k_B T$ for a cylinder oriented perpendicular to the cavity wall, and $1.01 k_B T$ for a cylinder oriented parallel to the cavity wall.
These numbers imply that the cylinder oriented parallel to the cavity wall ($\lambda = -0.5$) would be the most preferable configuration, which explains $\lambda=-0.5$ irrespective of the volume fraction $\phi$
and fraction of cylinders $\psi$.
At low volume fractions, both the spheres and cylinders are found in the bulk with cylinders oriented parallel to the cavity wall. However, as the volume fraction increases, cooperative effects related to the ordering of cylinders in the bulk lead the spheres to segregate to the cavity wall.
Also note that, even though we demonstrate the layering of particles in the density profiles and structure in the  orientational order parameter, the system is not crystalline; instead, it is still fluid-like, and particles diffuse throughout the system, as discussed in the following section.

\begin{figure}[!bt]
\centering
\includegraphics[width=0.98\columnwidth]{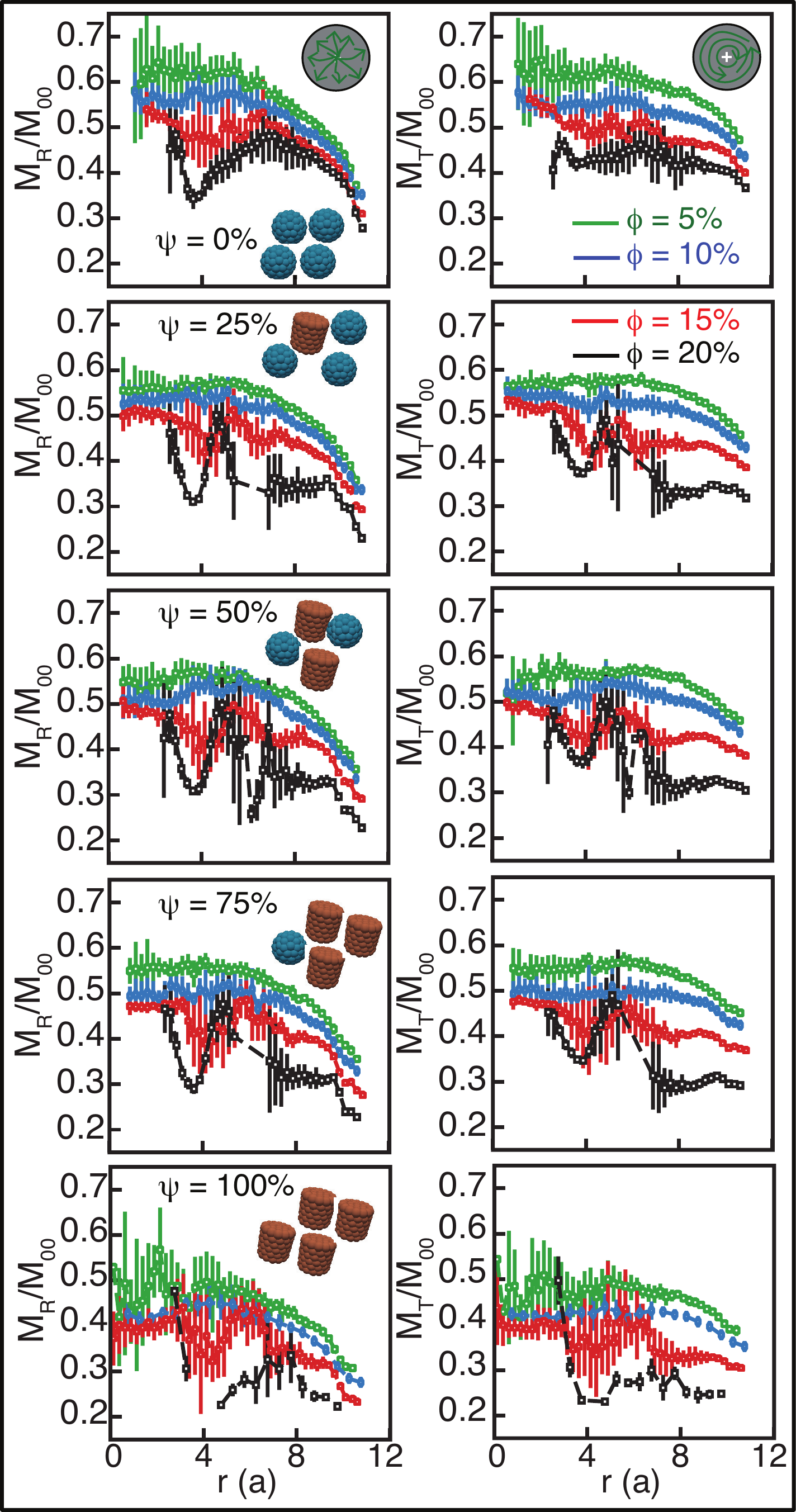}
\caption{
Radial mobility $M_R$ (left) and tangential mobility $M_T$ (right) for mixture of cylindrical particles with $r_C=2.62$ and $h_C=2r_C=5.24$, and spherical with $r_S=3$ confined in a
spherical cavity of $R=15$ for various particle concentrations, $\phi$, and different fraction of cylinders $\psi$.
Both components of mobility are normalized by the mobility of spherical particles in the bulk at infinite
dilution $M_{00}$ for $t \to 0$.
The error bars represent the statistical error. }
\label{fig:mobility}
\end{figure}

\subsection{Local mobility of the particles in the cavity.}
As mentioned earlier, recent experiments~\cite{carbajal2007asymmetry, eral2010anisotropic, Lara2011, hunter2014boundary} and simulations~\cite{zia:2016, zia2018} suggest that confinement can lead to anisotropic self--diffusion, which is not the case for unconfined suspensions.
To examine this, mobilities (short-time diffusivity) in both the radial and tangential directions are calculated using the Einstein--Stokes relation~\cite{Lara2011}
\begin{eqnarray}
\langle{\Delta \mathbf{x}_\text{R}^2(t)}\rangle(r_i) &=& 2M_\text{R}(r_i)t, \\
\langle{\Delta \mathbf{x}_\text{T}^2(t)}\rangle(r_i) &=& 4M_\text{T}(r_i)t,
\end{eqnarray}
for short--time $t \to 0$; $\Delta \mathbf{x} = \mathbf{x}(t+dt) - \mathbf{x}$(t),
$\Delta \mathbf{x}_\text{R}=\Delta\mathbf{x}\cdot\mathbf{x}/|\mathbf{x}|$, $\Delta \mathbf{x}_\text{T} = \Delta\mathbf{x} - \Delta\mathbf{x}_\text{R}$
denote the radial and tangential displacements, respectively.
$M_\text{R}(r_i)$ and $M_\text{T}(r_i)$ correspond to the instantaneous radial and tangential mobilities at radial location $r_i$ in an infinitesimal time interval $dt$.
Instantaneous radial and tangential mobilities are averaged in each shell during a simulation, and then over 10 independent realizations.

Figure~\ref{fig:mobility} displays both the radial $M_\text{R}$ and tangential $M_\text{T}$ components of mobility within the cavity for mixtures of spherical and cylindrical particles as a function of radial distance for various particle concentrations. Note that the two components are normalized by the mobility of a spherical particle at infinite dilution $M_{00}$.
A few observations can be drawn:
mobilities along both directions are not constant along the radial direction; instead, the particles diffuse fastest at the cavity
center and slowest at the cavity wall.
Second, both $M_\text{R}$ and $M_\text{T}$ decrease with increasing particle concentration due to enhanced many--body hydrodynamic interactions with $\phi$.
Next, the peaks and trough in mobility appear at the same radial position, corresponding to the local particle density as shown in Fig.~\ref{fig:density}, thereby revealing a correlation between structure and dynamics.
This becomes particularly apparent for the case of $\phi=20\%$, where a layered structure for both $M_\text{R}$ and $M_\text{T}$ corresponds to a similar density profile as observed in Fig.~\ref{fig:density}, e.g., the dip in mobility at $2<r<7$ corresponds to the peak in $\rho$ in the same radial range.
Both $M_\text{R}$ and $M_\text{T}$ display the expected decrease with increasing $\phi$ close to the wall for all cylinder fractions.
For $\phi=$ 5 and 10\%, we observe an expected decrease in mobility with $\phi$ in the bulk as well.
In contrast, for higher particle concentrations ($\phi$= 15\%), the mobility does not exhibit a decrease with $\phi$ in the bulk, and mobility for $\phi= 20\%$ at radial location $r \sim 5$ becomes equal or even larger than that for $\phi=15\%$.
Taken together, these observations reveal that a structural inhomogeneity leads to unexpected inhomogeneities in the corresponding mobility.
%

\subsection{Long time mobility of the particles.}
The displacement of a Brownian particle in a confined system is hindered,
and thus the mean square displacement (MSD) over time is lower than that observed in a bulk system~\cite{Chow:2015ja,zia2018}.
Our recent work on pure suspensions~\citep{unpublishedkeyA} showed that a change in shape from spherical to cylindrical at constant volume fraction leads to slower particle diffusion.
The question that arises here is: how does the fraction of cylinders in a mixture affect long-time dynamics?

\begin{figure*}[hbt]
\centering
\includegraphics[width=0.95\textwidth]{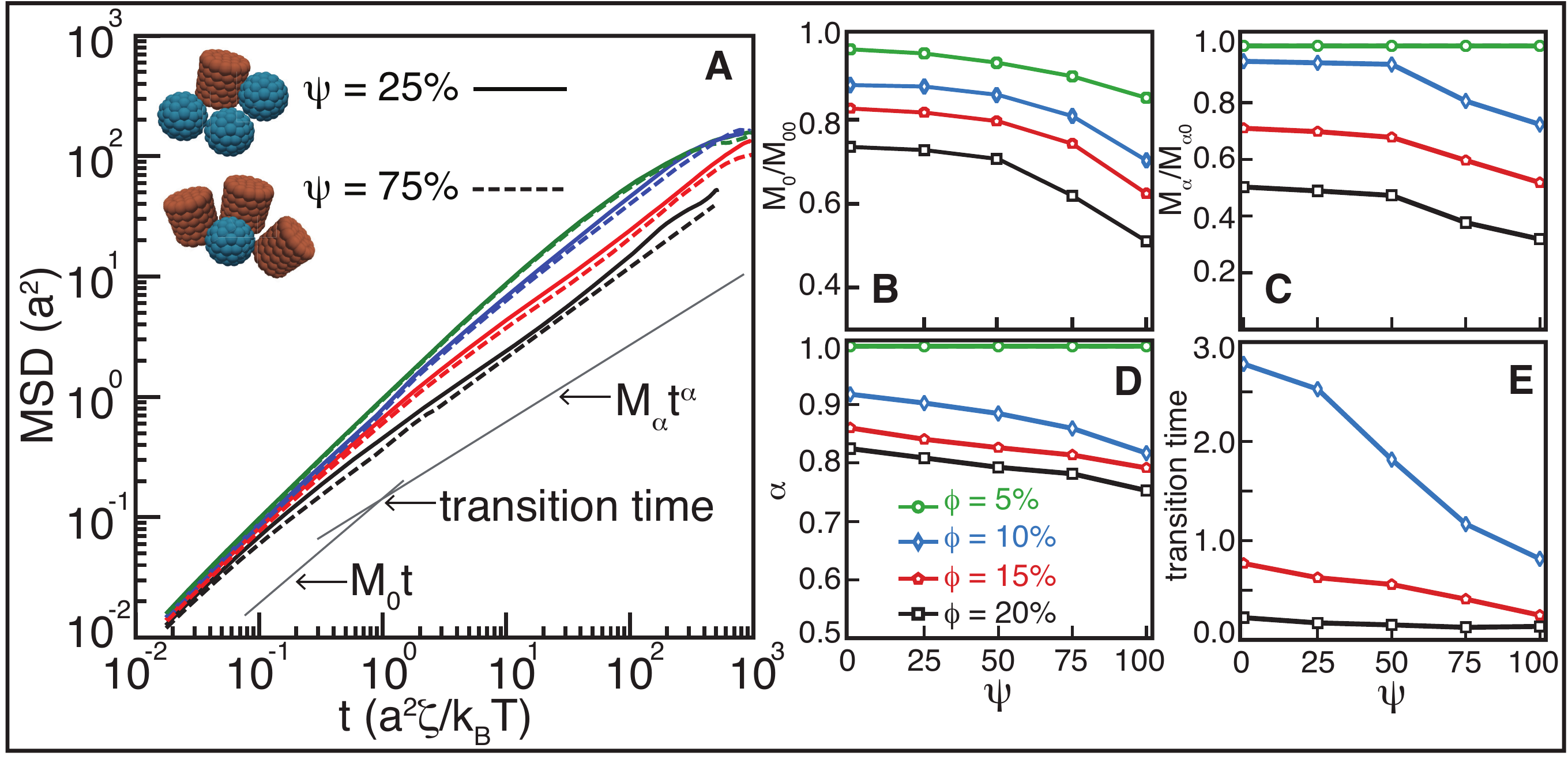}
\caption{
(A) Mean square displacement as a function of time for mixture of spherical and cylindrical particles that are confined in a spherical cavity with $R = 15$ for various particle concentrations $\phi$ with varying fraction of cylinders.
Solid and dashed lines are the results for $\psi=$ 25, and 75 \% fraction of cylinders.
(B) short time mobility scaled with the mobility of spherical particles in bulk at infinite
dilution $M_{00}$ plotted against $\psi$,
(C) sub-diffusive (at intermediate time) mobility scaled with the mobility of spherical particles in bulk at infinite
dilution $M_{00}$ plotted against $\psi$,
(D) sub-diffusive exponent $\alpha$ plotted against $\psi$,
and
(E) time scale to make transition between diffusive to sub--diffusive behavior plotted as a function of  fraction of cylinders for various values of $\phi$.
}
\label{fig:MSD}
\end{figure*}

Fig.~\ref{fig:MSD}A shows the average MSDs for mixtures at different particle concentrations.
The MSD of each system is calculated from ten independent simulations in which the particles diffuse for more than 300 particle diffusion times within the cavity.
A couple of observations can be drawn.
First, the MSDs grow linearly as short times, $t \to 0$, and reach a plateau in the long time limit, $t \to \infty$.
Second, at intermediate times, particle motion becomes sub--diffusive for systems with $\phi \ge 10\%$.

To analyze the diffusion behavior in detail we express the MSD as the generalized Stokes-Einstein relation
\begin{equation}\label{eq:MSD-time}
\big\langle \big(  \mathbf{R}(t)-\mathbf{R}(0)  \big)^2 \big\rangle = M_i t^i,
\end{equation}
where $\mathbf{R}$ is the $3N_p$ particle coordinate vector, $M_i$ is the generalized particle mobility coefficient and $i$ is the power law exponent that characterizes whether the particle motion is sub--diffusive $(\alpha<1)$, diffusive $(\alpha=1)$ or super-diffusive $(\alpha>1)$.
For the case of $(\alpha=1)$, the mobility $M_0$ is the diffusion coefficient.

Figure~\ref{fig:MSD}B--E displays our findings for the mean square displacement.
Short time mobility (Fig.~\ref{fig:MSD}B) and sub--diffusive mobility (Fig.~\ref{fig:MSD}C) decrease with increasing cylinder fraction $\psi$ and particle concentration $\phi$.
The dependence of $M_{0}/M_{00}$ and $M_{\alpha}/M_{00}$ on particle concentration $\phi$ is monotonic;
however, the dependence on cylinder fraction is weak at smaller cylinder fraction $\psi$ and becomes strong beyond 50\%, showing a smooth to "rigid" transition at $\psi=$ 50\%.
Note that for the case of $\phi = 5\%$, sub--diffusion is not observed at any cylinder fraction $\psi$.
On the other hand, the sub--diffusive--exponent $\alpha$ (Fig.~\ref{fig:MSD}D) that characterizes the strength of sub--diffusive behavior decreases with both $\phi$ and $\psi$.
In the case of short--time and sub--diffusive mobilities $M_{0}/M_{00}$ and $M_{\alpha}/M_{00}$, the dependence on $\psi$ is monotonic; however, the correlation gets less pronounced with increasing particle concentration. Finally, the transition time, defined as the time at which the system transitions from the short--time diffusive to the intermediate time sub--diffusive regime, is displayed in Fig.~\ref{fig:MSD}E.
We observe that for $\phi=0.1$ the transition time decreases strongly with the cylinder fraction and becomes nearly independent of cylinder fraction for higher particle concentrations.

To explain the smooth to rigid transition as observed in both $M_{0}/M_{00}$ and $M_{\alpha}/M_{00}$, we refer to the scaled density profiles with increasing $\psi$.
We observe that the scaled densities $n_T/\text{max}(n_T)$ for $\psi=0$ and 25\% are similar but then drop drastically for $\psi \ge$ 50\%, which affects the mobility in both the
diffusive and sub--diffusive cases.
On the other hand, for the transition time between the diffusive to the sub--diffusive regime (i.e. roughly the time needed for particles to diffuse a distance nearly equal to the radius), collisions between particles slow down their motion, leading to sub--diffusive behavior.
It follows that the transition time would decrease with increasing particle concentration $\phi$.
Further, at low $\phi$, changing shape from spheres ($\psi=0$) to ($\psi=100\%$) to cylinders, due their larger aspect ratio the latter should feel each other at shorter--time scales, compared to what is seen for spheres,
and hence yield a transition time that decreases with $\psi$.
At larger $\phi$, the system is so dense that even for different packings, i.e., sphere--sphere, sphere--cylinder and cylinder--cylinder cases, the particle interaction time scales become similar.

\section{Conclusions}\label{sec:conc}
We have studied the structure and dynamics of mixtures of finite size in mixtures of spherical and cylindrical particles confined in a spherical cavity. An Immersed Boundary-General geometry Ewald-like Method (IB-GgEm) approach was used in the corresponding calculations, thereby taking into account hydrodynamic interactions between particles and between particles and confining walls.
By systematically varying the cylinder fraction at different particle concentrations, it was found that particle shape has a pronounced effect on both the structure and dynamics
of confined Brownian suspensions.
Our results suggest that introducing non--spherical particles affects the local structure and local dynamics and global dynamics in different ways.
At a local level, particles are found to segregate based on shape, with cylinders adopting conformations with high orientational order.
At a global level, cylinders give rise to pronounced differences in the short-time mobility, the sub--diffusive behavior at intermediate timescales, and the transition time from diffusive to sub-diffusive behavior.
Particle concentration has a strong effect on such transitions.
To start with, the mobility shows a smooth to stiff transition at 50\% cylinder fraction for all particle concentrations and the sharpness of this transition increases with particle concentration.
Secondly, the sub-diffusive slope and sub--diffusive exponent show mixed features as a function of cylinder fraction. The slope shows a smooth to stiff transition similar to that of the short time mobility. The exponent decreases smoothly with increasing cylinder fraction. In both cases, the dependence on the cylinder fraction is insensitive to particle concentration.
Lastly, the transition time from the diffusive to sub--diffusive regime depends strongly on the cylinder fraction for low particle concentrations $(\phi=0.1)$; that dependence weakens with increasing particle concentration.

As an outlook, the role of aspect ratio and electrostatic interaction on particle mobility in confined mixtures will be considered in future that arise in a realistic cell environment.

\begin{acknowledgments}
We acknowledge the support by the Department of Energy, Basic Energy Sciences, Materials Research Division through the AMEWS EFRC Center. The models and codes employed for this work were developed with support from the Midwest Center for Computational Materials (MICCOM) and Award DOE-SC0008631 (KFF).
We gratefully acknowledge the computing resources provided on Bebop (Blues), a high-performance computing cluster operated by the Laboratory Computing Resource Center at Argonne National Laboratory, and the University of Chicago Research Computing Center.
\end{acknowledgments}
\bibliography{reference}

\end{document}